\documentclass[preprint,proceedings]{rmaa}

\newcommand{\D}{\discretionary{}{}{}}

\SetYear{2007}
\SetConfTitle{XII Reuni\'on Regional Latinoamericana de la UAI}

\title{Unveiling the phenomenon of Double Periodic Variables}

\author{
R.E. Mennickent\altaffilmark{1} and Z. Ko{\l}aczkowski\altaffilmark{1,2} }

\altaffiltext{1}{Departamento de F\'\i{}sica, Universidad de Concepci\'on, Casilla 160-C,
Concepci\'on,  Chile
(rmennick@astro-udec.cl). 
}
\altaffiltext{2}{Instytut Astronomiczny Uniwersytetu Wroc{\l}awskiego, Kopernika 11, 51--622 Wroc{\l}aw, Poland   (zibi\D{}@astro-udec.\D{}cl).}

\shortauthor{Mennickent \& Kolaczkowski}

\shorttitle{Double Periodic Variables}

\suppressfulladdresses

\listofauthors{R. E. Mennickent, Z. Ko{\l}aczkowski}
\indexauthor{Mennickent, R. E.}
\indexauthor{Ko{\l}aczkowski, Z.}

\abstract{In this paper we give a brief report of our recent research on Double Periodic Variables (DPVs), 
including the discovery of DPVs in the Galaxy and some insights on the nature of their long-cycle variability. }
\resumen {En este art\'\i{}culo damos un reporte breve de nuestra investigaci\'on sobre Variables de Per\'\i{}odo Doble (DPVs), incluyendo el descubrimiento de DPVs en nuestra galaxia y algunas ideas acerca de la naturaleza de su per\'\i{}odo  largo. }
\addkeyword{Binaries: Eclipsing}
\addkeyword{Stars: Early-type}
\addkeyword{Stars: Evolution}
\addkeyword{Stars: Mass-loss}

\begin{document}
\maketitle

Double Periodic Variables (DPVs) are blue stars characterized by a short periodicity (1-16 days) and a long periodicity (50-600 days) in their light curves. They were  discovered in the Magellanic Clouds after a search for Be stars in the OGLE variable star catalog (Mennickent et al. 2003) and now they amount to $\approx$ 110 in the MCs and about a dozen in our Galaxy (Fig. 1).  A key feature of DPVs is the relation between short and long period, hereafter called superperiod. While the short period reflects the orbital motion of a binary (Mennickent et al. 2005), the superperiod was an enigma until recently, when evidence was presented for an interpretation based on cycles of formation and dissipation of circumbinary discs (Mennickent et al.  2008; M08).  We have found a significant infrared excess for all DPVs with available 2MASS data. To explain the DPV phenomenon, we  have developed an  hypothesis. In our view, DPVs are low mass ratio, semi-detached, intermediate mass binaries in a stage of mass transfer, probably at the end of Case-A or Case-B mass exchange (M08). The B-type primary is rotating very fast while accretes matter from the later type donor in an Algol-like configuration. Matter is also ejected from the binary  system through the outer Lagrangian points  forming a reservoir of circumbinary gas. The superperiod could be produced by a variable mass ejection rate linked to the binary period. We have initiated an observing campaign to determine fundamental parameters of DPVs, mainly based on the study of the bright eclipsing systems, in order to shed light on the DPV evolution stage and their connection with classical Algol stars. In this program, the sample of Galactic DPVs HD\,50526, AU Mon, V\,393\,Sco and V356\,Sgr are being studied intensively, in optical and infrared wavelengths  (Fig.\,2).   We have identified, based on high-resolution spectroscopy of some of these systems, at least three spectral components: (1) the secondary star revealed in absorption lines characteristics of a B-A type spectrum,  (2) an optically thin H$\alpha$ emitting region  (we observe sometimes double peak emission and sometimes  a single emission at one of the wings of the Balmer absorptions),
and (3) an optically thick region forming variable and broad helium absorption lines that could be signatures of  the inner part of the disc and/or the primary star. 
Our preliminary analysis indicates a mass ratio  ($M_{2}/M_{1}$)  between 0.2--0.4 
for OGLE LMC--SC8--125836 (M08), OGLE LMC--SC13--156235 
and V\,393 Sco.


One of the hallmarks of DPVs is the correlation between orbital period and superperiod. This observational feature could be explained 
identifying the superperiod as the time needed for the disc, to reach the 3:1 resonance radius. We conjecture that in DPVs primary star does not accumulate gas, because it probably rotates at critical velocity. Then the transferred matter forms a decretion disc around the primary that starts expanding inside its Roche lobe. While longer is the orbital period, larger is this Roche lobe and longer the superperiod. 
When the disc reaches the 3:1 resonance radius ($r_{r}$) it starts to precess (Whitehurst \& King 1991, Lubow 1992; WK91 and L92 respectively) and matter starts to flow outside the binary system. In our view DPVs are semi-detached  binaries  with a low mass ratio allowing the formation of a circumprimary disc. 

We make some estimates using the theory of decretion discs. In this theory, the
disc expands due to viscous stress. The diffusion velocity at a radius $r$ is given by:\\

$v_{d}(r) \approx \frac{r}{t} \approx \alpha \beta \sqrt{GM_{1}}r^{\frac{2\gamma-3}{2}}$\\

\noindent
(Tutukov \& Pavlyuchenkov 2004), where $\gamma$ is a parameter related to the mean free paths of turbulent elements and $\alpha$ and $\beta$ are free parameters describing a diffusive disc. The time needed for a mass element to drift from the star to the 3:1 resonance  radius ($r_{r}$) is identified with the superperiod and it is given at order of magnitude by:\\

$P_{long} \sim \frac{r_{r}}{v_{d}}   = \eta r_{r}^{\frac{5-2\gamma}{2}}  \approx  \eta (0.46 a)^{\frac{5-2\gamma}{2}}$ \\

\noindent
where $\eta = 1/(\alpha \beta \sqrt{GM_{1}})$ and we have used the approximation $r_{r} \approx 0.46a$, where $a$ is the binary separation (WK91).\\



In the case $\gamma$ = 1, i.e. when the mean free paths of turbulent elements in the disc are proportional to the distance from the central gravitational object (and also proportional to the thickness of the disc, Tutukov \& Pavlyuchenkov 2004), we get:\\

$P_{long}  \sim 0.31 \eta a^{3/2}  \sim   \frac{0.3}{2\pi \alpha \beta} P_{orb}$\\

\noindent
this expression gives the right linear dependence with the orbital period ($P_{orb}$).
The proportionality factor depends on the viscosity properties of the disc, parametrized in $\alpha$ and $\beta$ in the Tutukov \& Pavlyuchenkov 's picture. We need $\alpha \beta \sim 1.5\times10^{-3}$  to reproduce the order of magnitude of the observed DPV correlation. Incidentally, 
this number is of the same order of magnitude that the figure found for accretion discs of cataclysmic variables during their hot states (Tutukov \& Pavlyuchenkov 2004).

\begin{figure}[!t]
\includegraphics[angle=0,width=\columnwidth]{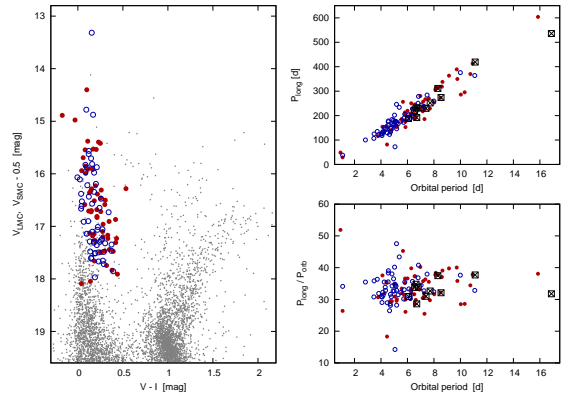}
\caption{Left panel: CMD of 59 LMC DPVs shown as
filled brown circles, open circles are used for 55 SMC DPVs. Additional background
(small dots) is a random sample of stars in the LMC.
We have adopted a distance moduli shift  -0.5 mag  for SMC
members. Right panel:
Graphs showing the $P_{orb}$-$P_{long}$ relationship for galactic (open squares) and MC DPVs.
Note the differences between the populations of the three different galaxies (Ko{\l}aczkowski \& Mennickent, in preparation).}

\label{fig:fig1}
\end{figure}

\begin{figure}[!t]
\includegraphics[angle=0,width=\columnwidth]{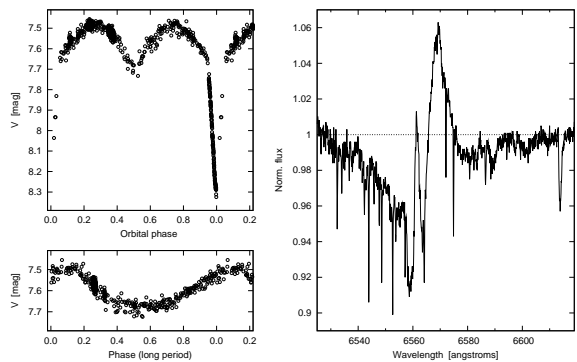}
\caption{Selected data for the galactic DPV V\,393 Sco.
Photometric variability of both periods $P_{orb}$= 7.7126 d  (left up) and
$P_{long}$= 252.0 d (left down) is shown based on ASAS-3 data 
(Pilecki and Szczygiel 2007). Part of one high resolution spectrum showing the H$\alpha$ line region
(right, Mennickent et al. in preparation).}
\label{fig:fig2}
\end{figure}

\end{document}